\documentclass[aps,pra,twocolumn,showpacs,amsmath,amsfonts,amssymb,floatfix,superscriptaddress]{revtex4}

\usepackage{bbold}									
\usepackage{color}
\usepackage{float}
\usepackage{graphicx}

\begin{document}


\newcommand{\diff}{\mathrm{d}}
\newcommand{\imagi}{\mathrm{i}}							
\newcommand{\br}{\mathbf{r}}							
\newcommand{\sumN}{\sum\limits_{i=1}^N }
\newcommand{\bra}{\langle}
\newcommand{\ket}{\rangle}


\title{Exact Density-Functionals with Initial-State Dependence and Memory}

\author{M. Ruggenthaler}
\affiliation{Department of Physics, Nanoscience Center, University of Jyv\"askyl\"a, 40014 Jyv\"askyl\"a, Finland}
\affiliation{Institut f\"ur Theoretische Physik, Universit\"at Innsbruck, Technikerstra{\ss}e 25, 6020 Innsbruck, Austria}
\author{S. E. B. Nielsen}
\affiliation{Lundbeck Center for Theoretical Chemistry, Department of Chemistry, Aarhus University, 8000 Aarhus C, Denmark}
\affiliation{qLEAP Center for Theoretical Chemistry, Department of Chemistry, Aarhus University, 8000 Aarhus C, Denmark}
\author{R. van Leeuwen}
\affiliation{Department of Physics, Nanoscience Center, University of Jyv\"askyl\"a, 40014 Jyv\"askyl\"a, Finland}
\affiliation{European Theoretical Spectroscopy Facility (ETSF)}

\date{\today}

\begin{abstract}
We analytically construct the wave function that, for a given initial state,
produces a prescribed density for a quantum ring with two non-interacting particles in a singlet state.
In this case the initial state is completely determined by the initial density,
the initial time-derivative of the density and a single integer that characterizes the (angular) momentum of the system.
We then give an exact analytic expression for the exchange-correlation potential
that relates two non-interacting systems with different initial states.
This is used to demonstrate how the Kohn-Sham procedure predicts the density of a reference system
without the need of solving the reference system's Schr\"odinger equation.
We further numerically construct the exchange-correlation potential for an analytically solvable system
of two electrons on a quantum ring with a squared cosine two-body interaction.
For the same case we derive an explicit analytic expression for the exchange-correlation kernel and analyze its frequency-dependence (memory) in detail.
We compare the result to simple adiabatic approximations and investigate the single-pole approximation.
These approximations fail to describe the doubly-excited states, but perform well in describing the singly-excited states.
\end{abstract}


\maketitle


\section{Introduction}

Time-dependent density-functional theory (TDDFT) \cite{CarstenBook,TDDFT}
allows for an exact description of a many-body system in terms of an effective non-interacting system,
known as the Kohn-Sham (KS) system.
The external potential (known as the KS potential) in the non-interacting system
is a functional of the density in such a way that the KS system has exactly the same density as the reference system.

The essential component in the KS construction is the exchange-correlation (xc) potential
that contains all non-trivial many-body effects. It depends on the initial states of the interacting
and the KS system (initial-state dependence) as well as the density at all previous times (memory).
Both features of the xc potential are, however, not well understood
and consequently virtually all commonly used approximations neglect them,
which in important cases (doubly-excited states, molecular dissociation, charge transfer etc.)
can lead to large errors in the calculated properties \cite{CarstenBook,TDDFT}.
It is therefore highly desirable to have exact analytical functionals available for model systems
that can serve as benchmarks and which can provide insight into how memory and initial state dependence
can be incorporated into approximate functionals for real systems.

In this work we explicitly construct such exact analytic functionals that do incorporate initial-state dependence and memory
for the case of a quantum ring (QR) with two particles in a singlet state.
In Sec.~\ref{sec:NonInt} we will derive functionals with an explicit initial-state dependence for the case that the two particles are non-interacting. These functionals will then be used to construct an explicit expression for the xc potential that connects two non-interacting systems. In Sec.~\ref{sec:Int} we will 
calculate the xc potential for two interacting particles at a specific density. For the same system we will then analytically construct the exact xc kernel of linear-response TDDFT and investigate its frequency-dependence. We conclude in Sec.~\ref{sec:Conclusion}.


\section{Functionals with Initial-State Dependence: Non-Interacting Model System} 
\label{sec:NonInt}

The dynamical properties of many-electron systems, such as molecules or solids is well-described by the
solution of the time-dependent Schr\"odinger equation (TDSE). If we restrict ourselves to external scalar potentials (such as laser fields in the dipole
approximation) then 
 the physical properties of an $N$-electron system evolving from a given initial state $|\Psi_0 \ket$ under the influence of an external scalar potential $v(\br,t)$ is determined by the Hamiltonian
\begin{equation}
\label{Ham}
\hat H(t) = - \frac{1}{2}\sumN \nabla_i^2 + \sum\limits_{i>j=1}^N w (|\br_i - \br_j|) + \sumN v(\br_i, t),
\end{equation}
where $\nabla_i$ is the gradient with respect to the spatial coordinate $\br_i$ and $w(|\br_i - \br_j|)$ is the electron-electron interaction (usually chosen to be Coulombic). In molecules and solids the form of the kinetic energy operator and the two-body interactions is always the same, whereas the external potential
$v$ varies from system to system. For this reason we will treat $v$ as a variable.
Consequently the solutions of the corresponding time-dependent Schr\"odinger equation (TDSE)
\begin{equation*}
\imagi \partial_t | \Psi(t) \ket  = \hat H(t) | \Psi(t) \ket 
\end{equation*}
can be uniquely labeled by the initial state and the external potential, i.e. the quantum states $| \Psi([\Psi_0,v],t) \ket$ depend functionally on the initial state and the external potential
\footnote{Actually, if we restrict the particles to a finite volume, the wave functions also depend on the chosen boundary conditions that make the Hamiltonian self-adjoint.}. 
However, due to the large number of degrees of freedom of the many-body wave function, a numerical solution of the TDSE is only feasible for small systems. 
In the KS approach of density-functional theory the interacting many-body problem is mapped onto an effective non-interacting system which considerably
reduces the computational effort. The effective potential in these equations is a functional of the density of the system which is defined as
\begin{equation*}
 n([\Psi_0,v],\br, t) = \bra \Psi([\Psi_0,v],t) | \hat n(\br) |Ê\Psi([\Psi_0,v],t) \ket,
\end{equation*}
where 
\begin{equation*}
\hat n(\br) = \sumN \delta (\br - \br_i),
\end{equation*}
is the density operator. 
A number of observables of large interest, such as the optical absorption spectrum in linear response or the time-dependent dipole moment, are
explicitly known as functionals of the density.
The basic theorems of DFT actually guarantee that, at least in principle, all observables are a functional of the density. This is a consequence of the fact that the Runge-Gross (RG) theorem \cite{RungeGross} and its generalizations \cite{GFPP, GFPP1D} guarantee (under certain assumptions) that the full many-body wave function is uniquely determined by only knowing its initial state and the density, i.e. the wave function is a functional of the initial state and the density 
$| \Psi([\Psi_0,n],t) \ket$. As a consequence the knowledge of $n[\Psi_0,v]$ is enough to calculate all physical properties of a many-body system. 

In Sec.~\ref{WaveFunc} we will give an explicit example of this result by analytically constructing the wave-function functional $| \Psi([\Psi_0,n],t) \ket$ for a specific system. How the density and the initial state determine the external potential of the Hamiltonian of Eq.~(\ref{Ham}) is then demonstrated in Sec.~\ref{PotFunc}. This result is then employed to give an example of the KS scheme, which can be used to predict the density of a reference system by solving an auxiliary non-interacting problem, by explicitly constructing an initial-state dependent xc potential in Sec.~\ref{XCFunc}. 

											
\subsection{Wave-Function Functional}							
\label{WaveFunc}

In this Section we give a non-trivial analytical realization of the wave-function functional $|\Psi([\Psi_0,n],t) \ket$  
with explicit initial-state dependence, for the case of two non-interacting particles on a QR
(a one-dimensional system with periodic boundary conditions) of length $L$.  

We assume the non-interacting wave function $| \Phi(t) \ket$ to be in a spin-singlet configuration.
In a position-spin basis we then make an orbital product Ansatz for the spatial part of the resulting wave function
\begin{equation} 
\label{Ansatz}
\Phi(x,y,t) = \varphi(x,t) \varphi(y,t) ,
\end{equation}
where $x$ and $y$ are the spatial coordinates of the particles along the ring.
The full position-spin dependence is obtained by multiplication with the usual anti-symmetric singlet spin-function.
Here the orbital $\varphi(x,t)$ satisfies the one-dimensional Schr\"odinger equation		
\begin{equation} 
\label{Schroedinger}
\imagi \partial_t \varphi(x,t) = \left( -\tfrac{1}{2} \partial_x^2  + v_s(x,t) \right) \varphi(x,t) ,
\end{equation}
with periodic boundary conditions on the interval $[0,L]$ and starting from the initial state $\varphi_0(x)=\varphi(x,t_0)$ (We adopt the convention that an external potential belongs to a non-interacting system if we use the subindex $s$).
We may now rewrite the orbital in terms of real-valued functions $|\varphi|$ and $S$ as \cite{Vignale1999,Maitra2001,CarstenBook}
\begin{equation*}
\varphi(x,t)= |\varphi(x,t)| \exp(\imagi S(x,t)) .
\end{equation*}
The periodic boundary conditions on the orbital $\varphi(x,t)$ then correspond to
periodic boundary conditions on the norm $|\varphi|$ and quasi-periodic boundary conditions on the phase $S$, i.e.
\begin{align}
\label{period1}
S(L,t) &= S(0,t) + 2 \pi m , \\
\label{period2}
\partial_x S(L,t) &= \partial_x S(0,t) ,
\end{align}
for some integer $m$. Note that the initial orbital $\varphi_0(x) = |\varphi_0(x)| \exp(\imagi S_0(x))$
determines the choice of $m$ since $S_0 (x)= S(x,t_0)$ must obey condition (\ref{period1}).
To proceed, we use that the density and current of the non-interacting system,
\begin{align}
n(x,t) & = 2 |\varphi(x,t)|^2 , \\
\label{current}
j(x,t) & =  n(x,t) \partial_x S(x, t) ,
\end{align}
are connected by the continuity equation
\begin{equation} 
\label{continuity}
- \partial_x j(x, t) = - \partial_x \left[ n(x,t) \partial_x S(x, t)\right] = \partial_t n(x,t) ,
\end{equation}
which expresses the local conservation of particles.
This is a Sturm-Liouville equation \cite{GFPP1D} depending parametrically on the time $t$ and
thus the density determines the phase function $S(x,t)$ for a given set of boundary conditions (\ref{period1}) and (\ref{period2}).
More precisely $S$ is determined uniquely up to a purely time-dependent constant $C(t)$,
since the constant function is eigenfunction of the Sturm-Liouville operator in Eq.~(\ref{continuity})
with eigenvalue zero and also satisfies the boundary conditions (\ref{period1}) and (\ref{period2}).
Physically this freedom amounts to the gauge freedom in the potential.
Following similar derivations as in reference \cite{GFPP1D} we find that
\begin{align} 
\label{Thetam} 
S([m,n],x,t) &= \int_0^L \diff y \, K_t(x,y) \partial_t n(y,t)  \\
&+ \frac{2 \pi m}{\int_0^L \frac{\diff z}{n(z, t)}} \int_0^x \frac{\diff z}{n(z,t)} + C (t), \nonumber
\end{align}
where we defined
\begin{align*}
K_t([n],x,y) &= \frac{1}{2} [ \theta (y-x) - \theta (x-y) ] \int_y^x \frac{\diff z}{n (z, t)} \\
&- \frac{\eta(x,t) \eta(y,t)}{\int_0^L \frac{\diff y}{n(y,t)}} ,
\end{align*}
with $\theta$ the Heaviside function and
\begin{equation*}
\eta (xt) = \frac{1}{2} \left( \int_0^x \frac{\diff y}{ n (y, t)} + \int_L^x\frac{ \diff y}{ n (y, t)}\right) .
\end{equation*}
Note, these functions are defined only within the interval $[0,L]$ but can be extended periodically outside of it.
At $t=t_0$ this equation determines $S_0 (x)$ in terms of $n(x,t_0)$, $\partial_t n(x,t_0)$ and $m$,
up to an overall constant and therefore for a given choice of $m$ the density completely determines the initial state $\varphi_0 (x)$
up to a global phase factor $e^{\imagi \alpha}$.
Thus, if we restrict ourselves to the product Ansatz of Eq.~(\ref{Ansatz}),
there is only a countably infinite number of physically different initial states possible for any given time-dependent density.
Obviously these initial states all share the same initial density and time-derivative of the density, but their phases differ.  If we compare the resulting currents given by Eq.~(\ref{current}) as functionals of the density and the initial state $m$ we find with the help of Eq.~(\ref{Thetam}) that
\begin{equation*}
 j([m,n],x,t) - j([m',n],x,t) = \frac{2 \pi (m-m')}{\int_0^L \frac{\diff x}{n(x, t)}},
\end{equation*}
i.e. the currents differ only by a time-dependent constant. Accordingly the integral of the local velocity fields $v(x,t) = j(x,t)/n(x,t)$ differ exactly by 
$2 \pi (m-m')$.
So the density rotates differently around the QR for the different values of $m$, but in such a way as to yield the same density.

The resulting density-functional for the orbital (and with this the full wave function) is then given by
\begin{align*}
\varphi([m,n],x,t) = \sqrt{\frac{n(x,t)}{2}}\exp\left(\imagi S([m,n],x,t)\right).
\end{align*}
This is an explicit realization of the RG result. As pointed out before, a direct consequence is that we can calculate all observables of the particles in terms of the density and the choice of initial state only.
For instance, the kinetic-energy functional in this case becomes	
\begin{align*} 					 
T([m,n],t) = &\frac{1}{8} \int_0^L \diff x \, \frac{\left(\partial_x n(x,t)\right)^2}{n(x,t)}
\\
&+ \frac{1}{2}\int_0^L \diff x\,n(x,t)\left(\partial_x S([m,n],x,t) \right)^2 , \nonumber
\end{align*}
where the first term on the right hand side is the famous Weizs\"acker kinetic-energy functional. The second term is an initial-state dependent correction that together with the Weizs\"acker term constitutes the exact kinetic energy-functional. 


\subsection{Potential Functional}
\label{PotFunc}

The basic theorems of TDDFT further establish the uniqueness and existence of a density-potential mapping, i.e. for a given initial state there is a one-to-one correspondence between the external potentials and the densities. This allows for the determination of the external potential that produces a given density by propagation of an initial state, i.e. the external potential is a functional of the initial state and the density $v([\Psi_0,n],\br,t)$. This fact forms the basis of the KS construction, which allows us to determine the density of an interacting system by solving an auxiliary non-interacting problem.
\\

Here we will give an explicit example for the functional $v[\Psi_0,n]$. We will rely upon our previous results of two non-interacting particles on a QR.
The external potential $v_s$ can readily be expressed in terms of the orbital by inverting the Schr\"odinger Eq.~(\ref{Schroedinger}) and we find \cite{Vignale1999,Maitra2001,CarstenBook}
\begin{align*} 
v_s&([\varphi],x,t) = \frac{\imagi \partial_t \varphi(x,t)+ \frac{1}{2} \partial_x^2\varphi(x,t)}{\varphi(x,t)} \\
&=\frac{1}{2} \frac{\partial_x^2 |\varphi(x,t)|}{|\varphi(x,t)|} -\partial_t S(x,t) - \frac{1}{2}\left(  \partial_x S(x,t)\right)^2 \nonumber
\\
&  +\imagi \left\{ \frac{\partial_t |\varphi(x,t)|}{|\varphi(x,t)|} + \frac{\partial_x |\varphi(x,t)|}{|\varphi(x,t)|}  \partial_x S(x,t) + \frac{1}{2}  \partial_x^2 S(x,t) \right\} . \nonumber
\end{align*}
The last term on the right hand side vanishes as a consequence of the continuity Eq.~(\ref{continuity}) and we thus find using $|\varphi| = \sqrt{n/2}$ that
\begin{align} 
\label{vfunctional} 
v_s([m,n], x, t) &= \frac{1}{2} \frac{\partial_x^2 \sqrt{n(x,t)}}{\sqrt{n(x,t)}}  - \partial_t S([m,n],x,t) \\
&- \frac{1}{2}\left( \partial_x S([m,n],x,t) \right)^2 , \nonumber
\end{align}
which gives $v_s$ as a functional of $n$ and the initial state (characterized by $m$).
The potential $v_s[ m,n]$ exists whenever we have a unique $ S[m,n]$, i.e. for $n>0$ and the integrability conditions $\int_0^L \diff x \left|1/ n(x,t) \right| < \infty$ and $ \int_0^L \diff x | \partial_t n(x,t) | < \infty$ are fulfilled \cite{GFPP1D}.
Thus we have analytically defined a density-potential mapping which is also explicitly initial-state dependent.
We stress that the periodic boundary conditions on the wave function were essential in deriving Eq.~(\ref{vfunctional}).
This excludes, for instance, the example of a homogeneous electric field on a ring of constant density given in reference 
\cite{MaitraPBC}
\footnote{If we take $|\varphi| = \sqrt{n/2}$ to be constant and $S(x,t)=2\pi x j /L  -x \int_0^t \diff t' \mathcal{E} (t') + c(t)$ with $j$ integer, then Eq.~(\ref{vfunctional}) yields $v= x \mathcal{E}(t)$ for appropriately chosen $c(t)$.
This potential describes a homogeneous electric field. However, we see that this choice of
$S$ violates the condition (\ref{period1}) (which should be valid for all times) unless $\mathcal{E}(t) = 0$.}.


\subsection{Exchange-Correlation Functional}
\label{XCFunc}

The functional $v[\Psi_0,n]$ plays a central role in TDDFT. 
In practice, however, we are usually not directly interested in this mapping. We are
rather interested in the density of a particular system that has a specific external potential $v_{\rm ext}$. For example, in the case that
we want to describe a single molecule in a laser field, the potential $v_{\rm ext}$ is simply given by the Coulombic attraction of the atomic nuclei in the molecule with the addition of the laser field. For a given choice of $v_{\rm ext}$ 
every observable we want to know is then determined by solution of the TDSE for the given initial state $| \Psi_0 \rangle$. 
In particular we can calculate the density of the system, which, for future reference, we denote by $n_0 (\br, t)$. 
However, the full solution of the TDSE is usually not feasible in practice, due to the large
degrees of freedom that we need to consider. The main idea of the KS construction in TDDFT is to reduce the complexity by mapping the
interacting many-body problem to a non-interacting many-body problem with the same density. This leads to one-particle equations that are computationally much
easier to deal with. The price we pay for this simplification is that the functional $v[\Psi_0,n]$ now appears implicitly as part of the
xc potential $v_{\mathrm{xc}}$ in the KS equations. Below we will present an analytic example of an xc potential
for our QR system. However, we will start with a brief description of the KS method and define the KS and xc potentials.

The existence of a density-potential mapping $v[\Psi_0,n]$ does not depend on the chosen two-body interaction. Specifically this means that we have a density-potential mapping for interacting as well as non-interacting systems. For the case of a non-interacting system this mapping is called $v_s [\Phi_0, n]$.
Since in this case we have no two-body interactions the Hamiltonian is then simply given by
\begin{equation*}
\hat H_s(t) = - \frac{1}{2}\sumN \nabla_i^2 + \sumN v_s(\br_i, t).
\end{equation*}
The initial state $|\Phi_0 \ket$ of the non-interacting system is usually chosen to be a single Slater determinant of orbitals $\varphi_i(\br)$. This allows us 
to reduce the TDSE for the non-interacting system of $2 N$ electrons to single-orbital equations of the form
\begin{eqnarray} 
\label{TDKSEq}
\imagi \partial_t \varphi_i(\br, t) &=& \left[ -\tfrac{1}{2} \nabla^2 + v_s ([\Phi_0,n],\br, t) \right] \varphi_i(\br, t) ,\\
n (\br, t) &=& 2 \sum_{j=1}^{N} |\varphi_j (\br, t)|^2
\label{TDKSEq2}
\end{eqnarray}
where $\varphi_i (\br, t_0)=\varphi_i (\br)$.
By definition of the functional $v_s [\Phi_0,n]$ \cite{CarstenBook, TDDFT}, a density $n(\br,t)$ compatible with $n(\br,t_0)=\langle \Phi_0 | \hat{n} (\br) | \Phi_0 \rangle$ \footnote{To be precise, also the first time-derivative of the density $n(\br,t)$ at the initial time $t=t_0$ has to be compatible with the chosen initial state due to the continuity equation.} can be reconstructed from solving Eqs.~(\ref{TDKSEq}) and (\ref{TDKSEq2}). 
In particular, if $n=n_0$ is the density of an interacting system with
external potential $v_{\rm ext}$ and initial state $| \Psi_0 \rangle$ then, provided that we chose $| \Phi_0 \rangle$ such that
$\langle \Phi_0 |\hat{n} (\br)| \Phi_0 \rangle = \langle \Psi_0 |\hat{n} (\br)| \Psi_0 \rangle$, the potential $v_s [\Phi_0,n_0]$
reproduces the density $n_0$ of the interacting system in a non-interacting system. 
However, it is clear that the Eqs.~(\ref{TDKSEq}) and (\ref{TDKSEq2}) can not predict the density $n_0 (\br, t)$ of interest since
they contain no information on the interacting system that we are trying to solve.
 To set up a predictive scheme we need to connect the interacting and the non-interacting system.
To do this we introduce the KS potential \footnote{We point out, that in the literature the term ``KS potential'' is often also used to refer to the different potentials $v_s$ irrespective of their functional dependence. Here we employ this term exclusively to the functional defined in (\ref{vKS}).}
\begin{equation} 
\label{vKS}
v_{\mathrm{KS}}[\Psi_0,\Phi_0,n,v_{\mathrm{ext}}] = v_{\mathrm{ext}} + v_s[\Phi_0,n] - v[\Psi_0,n].
\end{equation}
If we assume full knowledge of the functionals $v[\Psi_0,n]$ and $v_s[\Phi_0,n]$ then the set of equations
\begin{eqnarray} 
\label{TDKSEq3}
\imagi \partial_t \varphi_i(\br, t) &=& \left[ -\tfrac{1}{2} \nabla^2 + v_{\rm KS} (\br, t) \right] \varphi_i(\br, t) , \\
n (\br, t) &=& 2 \sum_{j=1}^{N} |\varphi_j (\br, t)|^2 
\label{TDKSEq4}
\end{eqnarray}
does have a unique solution \cite{vanLeeuwen,GFPP, GFPP1D} for a self-consistent density $n_{\mathrm{sc}}$. 
By definition of $v_s [\Phi_0,n]$ the self-consistent density $n_{\mathrm{sc}}$ is exactly attained whenever
\begin{equation*}
v_{\mathrm{KS}}[\Psi_0,\Phi_0,n_{\mathrm{sc}},v_{\mathrm{ext}}] = v_s [\Phi_0, n_{\mathrm{sc}}], 
\end{equation*}
which according to Eq.~(\ref{vKS}) is precisely satisfied when
\begin{equation*}
v_{\mathrm{ext}} = v [\Psi_0, n_{\mathrm{sc}}].
\end{equation*}
In turn, this is exactly true when $n_{\mathrm{sc}}=n_0$ as there is a unique potential producing a given density. 
We therefore see that the set of Eqs.~(\ref{TDKSEq3}) and (\ref{TDKSEq4}) has exactly a self-consistent 
solution at the density $n_0$ of the interacting system with initial state $| \Psi_0 \rangle$ and external potential $v_{\mathrm{ext}}$.
To make the scheme practical we need to know the functional $v_s [\Phi_0,n] - v [\Psi_0,n]$ of Eq.~(\ref{vKS}) or at least have a reasonable
approximation for it.

The first non-trivial approximation to this expression is given by
the classical electrostatic potential of the electrons, i.e. the Hartree potential
\begin{equation*} 
v_\mathrm{H}([n],\br, t) = \int \diff^3 r' n(\br', t) w(|\br-\br'|).
\end{equation*}
Usually this approximation is made explicit and the rest is then called the xc potential $v_\mathrm{xc}[\Psi_0,\Phi_0,n]$,
\begin{equation*} 
v_s[\Phi_0,n] - v[\Psi_0,n] = v_\mathrm{H}[n] + v_\mathrm{xc}[\Psi_0,\Phi_0,n] .
\end{equation*}
The KS potential may thus also be written as
\begin{equation*}
v_\mathrm{KS}[\Psi_0,\Phi_0,n,v_{\mathrm{ext}}] = v_{\mathrm{ext}} + v_\mathrm{H}[n] + v_\mathrm{xc}[\Psi_0,\Phi_0,n] .
\end{equation*}
Therefore, the fundamental approximation in TDDFT is that of the xc potential
$v_\mathrm{xc}[\Psi_0,\Phi_0,n]$ and the results thus only depend on the quality of this approximation.

However, the xc potential $v_\mathrm{xc}[\Psi_0,\Phi_0,n]$ is still a complicated functional
that depends on the initial states of both the interacting and non-interacting system (initial-state dependence)
and the density at all previous times (memory).

Let us now give an example for the KS and xc potentials for our model system.
The construction of these functionals requires the knowledge of the functional $v[\Psi_0,n]$, which is not explicitly known. However, if the reference system is also non-interacting then $v[\Psi_0,n]=v_s [\Psi_0,n]$ and $v_\mathrm{H}[n]=0$, and we find that					
\begin{align*}					
v_\mathrm{KS}[\Psi_0,\Phi_0,n,v_{\mathrm{ext}}] &= v_{\mathrm{ext}} + v_s[\Phi_0,n] - v_s[\Psi_0,n] , \\
v_\mathrm{xc} [\Psi_0,\Phi_0,n] &= v_s [\Phi_0,n]  - v_s [\Psi_0,n] .
\end{align*}
For our case of a QR with two particles in a single-orbital singlet state the functional $v_s [\Phi_0,n]$ is given by Eq.~(\ref{vfunctional}), and we find			
\begin{align*} 
 &v_\mathrm{xc}([m,m',n],x,t)  = v_s ([m',n],x,t)-v_s ([m,n],x,t) \nonumber \\
 & = 2 \pi (m-m') \; \partial_t \left(\frac{ \int_0^x \frac{\diff z}{n(z,t)}}{\int_{0}^{L} \frac{\diff z}{ n(z, t)}}\right) \nonumber
  + \frac{2 \pi^2 (m^2- m'^2)}{\left( n(x,t) \,\int_{0}^{L} \frac{\diff z}{ n(z, t)}\right)^2} \nonumber \\
 & + \frac{2 \pi (m- m')}{\int_{0}^{L} \frac{\diff z}{ n(z, t)}} \frac{ \partial_x S([ 0,n], x,t)}{n(x t)} .
\end{align*}
where $\partial_x S([0,n], x,t)$ is defined only in terms of $n$ and $\partial_t n$ and corresponds to 
the spatial derivative of the first term on the right hand side of Eq.~(\ref{Thetam}).
Note, the integers $m$ and $m'$ play the role of the initial state $| \Psi_0 \rangle$ respectively $| \Phi_0 \rangle$.
The corresponding KS equations are thus
\begin{align*} 
\imagi \partial_t \varphi (x,t) &= \Big( -\tfrac{1}{2} \partial_x^2  + v_{\rm{ext}}(x,t)  \nonumber \\
&+ v_\mathrm{xc} ([m,m',n],x,t) \Big) \varphi (x,t), 
\\
n(x,t) &= 2 |\varphi (x,t)|^2, 
\end{align*}
with $\varphi (x,t_0)=\varphi_0^{m'}(x)$.
This equation determines the density $n(x,t)$ of the reference system when we prescribe $v_{\mathrm{ext}}$.
We note that the xc potential is given only in terms of $n$ and $\partial_t n$. 
In contrast, the functional $v_s([m,n],x, t)$ of Eq.~(\ref{vfunctional}) that reproduces a \textit{prescribed} density via propagation of the KS equation also contains a second-order time-derivative of the density (in the term $\partial_t S$ as can be seen with the help of Eq.~(\ref{Thetam})). We therefore can explicitly see that the second order time-derivative of the density vanishes if we connect the two systems. This is an important fact which sometimes is overlooked in the literature and can lead to misunderstandings about the KS approach \cite{Schirmer, Maitra, CarstenBook}.


\section{Functionals with Memory: Interacting Model System} 
\label{sec:Int}

In the previous section we have constructed functionals that depend only on the density at one time. Although also time-derivatives of the densities appear in the expressions we call these functionals time-local and accordingly they do not exhibit memory. At this point it is useful to give a more precise definition of memory. We first define the xc kernel as the functional derivative of
the xc potential, i.e.
\begin{equation}
f_{\mathrm{xc}} (\br, t, \br', t') = \frac{\delta v_{\mathrm{xc}} (\br, t)}{\delta n(\br', t')}  .
\label{xc-kernel-derv}
\end{equation}
Any approximation to the xc potential that depends only locally on the density and its time-derivatives gives rise to an xc kernel
that is proportional to time-derivatives of the delta function $\delta (t-t')$. These functions vanish for $t \neq t'$
and therefore have zero memory depth. If the xc kernel is non-zero for $t \neq t'$ we will say that the xc potential has memory. 
We can find another useful characterization of memory in the case that the functional derivative of Eq.~(\ref{xc-kernel-derv}) is evaluated
at a ground state density. Due to the time translation invariance of the
ground state Hamiltonian the kernel $f_{\mathrm{xc}}$ will then only depend on the time-arguments through the combination $t-t'$, i.e.
$f_{\mathrm{xc}} (\br, t, \br', t') = f_{\mathrm{xc}} (\br , \br',t- t')$. We can therefore by means of a Fourier transform define
a frequency-dependent xc kernel by
\begin{equation*}
f_{\mathrm{xc}} (\br,\br',\omega) = \int \diff \tau \, e^{\imagi \omega \tau} f_{\mathrm{xc}} (\br , \br', \tau).
\end{equation*}
In this case memory is characterized by a non-polynomial frequency dependence of $f_{\mathrm{xc}}$ (since the 
Fourier transform of the $n$-th time derivative of a delta function gives a frequency dependence proportional to $\omega^n$ ).

We now address the question whether for our QR system we can construct an xc potential with memory.
We have seen that the xc potential that arises in the modeling of a non-interacting system by another non-interacting system with a different
initial state has no memory (at least not for the product Ansatz used). One way to induce memory is to
introduce many-body interactions. 
However, in the case that the reference system is interacting we do not know $v_\mathrm{xc}[\Psi_0,\Phi_0,n]$
as we do not know $v[\Psi_0,n]$.
However, if we can determine $n[\Psi_0,v_{\mathrm{ext}}]$ for some $|\Psi_0\rangle$ and a specific external potential $v_{\mathrm{ext}}$,
we can still calculate $v_\mathrm{xc}[\Psi_0,\Phi_0,n,v_{\mathrm{ext}}]$ as a function of space and time for this density,
since $v[\Psi_0,n]$ is then known, i.e. $v[\Psi_0,n]=v_{\mathrm{ext}}$.
Here we will do this for the case of two particles on a QR of length $L$	    
with external potential $v_{\mathrm{ext}}=0$ and which interact via a squared cosine potential, i.e. for the Hamiltonian
\begin{equation} 
\label{ham1}
\hat{H} = -\frac{1}{2}\left( \partial_x^2 + \partial_y^2 \right)  + \lambda \cos^2\left(\frac{\pi}{L}(x-y)\right) ,
\end{equation}
where $\lambda$ is the strength of the interaction. In Sec.~\ref{sec:vxc} we will then construct the resulting xc potential which takes the simple form
\begin{equation}
\label{SpecificXCpot}
v_\mathrm{xc}=v_s[m',n] - v_\mathrm{H}[n],
\end{equation}
since in our example $v_{\mathrm{ext}}=0$. To give an explicit expression of functionals with memory we will further construct the xc kernel of TDDFT in Sec.~\ref{HxcKernel} for this system. The frequency-dependence (memory) of the xc kernel will then be investigated in detail in Sec.~\ref{sec:freq}. Finally, in Sec.~\ref{sec:SPA} we will test the validity of the single-pole approximation for this model system. However, to do all these things, it will prove helpful to first explicitly construct all the eigenstates of the Hamiltonian of Eq.~(\ref{ham1}).


\subsection{Spectrum of the Model System}
\label{sec:spec}

The eigenfunctions of the Hamiltonian of Eq.~(\ref{ham1}) can be written as the product of a spatial wave function $\Psi (x,y)$ and a spin-function. We have a spin-singlet (spin-triplet) configuration if $\Psi(x,y)$ is (anti)-symmetric with respect to an interchange of $x$ and $y$, i.e.
\begin{equation} 
\label{st}
\Psi (x,y) = \pm \Psi (y,x)
\end{equation}
where $+$ refers to the singlet state and $-$ to the triplet state. We further have the periodic boundary conditions
\begin{align*}
\Psi (x+L,y) &= \Psi (x,y) , \\
\Psi (x,y+L) &= \Psi (x,y) ,
\end{align*} 
with similar conditions on the spatial derivatives.
It is convenient to introduce the center-of-mass coordinate $R=(x+y)/2$ and the relative coordinate $r=x-y$.
In terms of these coordinates the Hamiltonian of Eq.~(\ref{ham1}) attains the form
\begin{equation*}
\hat{H} = -\frac{1}{4} \partial_R^2 -  \partial_r^2   + \lambda \cos^2\left(\frac{\pi r}{L} \right) .
\end{equation*}
The eigenstates $\Phi (R,r) = \Psi (x,y)$ in the new coordinates then satisfy the equivalent property of Eq.~(\ref{st})
\begin{equation} 
\label{sitri}
\Phi (R,r) = \pm \Phi (R,-r),
\end{equation}
and the periodic boundary conditions
\begin{equation} 
\label{boundary1}
\Phi (R+ \frac{L}{2}, r \pm L) = \Phi (R, r),
\end{equation}
and similarly for the spatial derivatives.
With the Ansatz  $\Phi(R,r)= f(R)g(r)$ the Schr\"odinger equation can be separated. The periodic boundary conditions on $f$ and $g$ become
\begin{align}
\label{gboundary}
g(r+L) &=\pm g(r) , \\
\label{fboundary}
f(R+L/2) &= \pm f(R) ,
\end{align}
and similarly for the spatial derivatives, where the signs on the right hand side of these equations must be the same for $f$ and $g$ in order to fulfill Eq.~(\ref{boundary1}).
The equation for the center-of-mass coordinate $R$ becomes a free particle Schr\"odinger equation
\begin{equation*}
- \tfrac{1}{4} \partial_R^2 f(R) = \epsilon f(R),
\end{equation*}
which has the eigenstates (up to normalization) 
\begin{equation*}
f(R) = \exp{\left(\frac{\imagi  2\pi k R }{L}\right)},
\end{equation*}
where the boundary conditions with $\pm$ in Eq.~(\ref{fboundary}) correspond to $k$ being
even and odd respectively. The energy eigenvalue is $\epsilon=(k \pi/L)^2$.
After changing coordinates to $z=r \pi/L $ the Schr\"odinger equation in the relative coordinate becomes
\begin{equation} 
\label{Mathieu}
\partial_z^2 M(z) +\left[ a - 2 q \cos\left(2 z \right)\right] M(z) = 0 ,
\end{equation}
where we defined $M(z)=g(L z/\pi)$. We further defined 
\begin{align*}
a &= \frac{L^2}{\pi^2}  \Big( E-\epsilon - \frac{\lambda}{2} \Big) , \\
q &= \frac{\lambda L^2}{4 \pi^2} ,
\end{align*}
with $E$ the eigenenergy of the full Hamiltonian of Eq.~(\ref{ham1}).
The boundary condition of Eq.~(\ref{gboundary}) then becomes $M(z+\pi)=\pm M(z)$.
Eq.~(\ref{Mathieu}) is the well-known Mathieu equation \cite{Stegun}. 
The solutions are given by the Mathieu-sine and Mathieu-cosine functions denoted by $SE (l,q,z)$ and $CE (l,q,z)$
where $l$ is a non-negative integer labelling certain discrete values $a_l$ for the constant $a$ in Eq.~(\ref{Mathieu}).
In the limit $\lambda \rightarrow 0$ (non-interacting case) we simply have
$CE(l, 0, z )= \cos(l z)$ and $SE(l,0,z)=\sin(l z)$ and $a_l=l^2$. We thus see that the $\pm$ signs in the boundary conditions
Eq.~(\ref{gboundary}) correspond to the case that $l$ is even and odd respectively.
From Eq.~(\ref{sitri}) we see that the singlet and triplet case corresponds to the symmetry
$g(r)=\pm g(-r)$ or equivalently $M(z)=\pm M(-z)$ for the Mathieu functions. This means that
the singlet solution corresponds to the Mathieu-cosine function and the triplet to the Mathieu-sine function.
The full solution of the problem is therefore given by
\begin{align*}
\Psi^{+}_{kl}(x,y) &= N_l^+ \exp\left(\frac{\imagi \pi}{L} k (x+y) \right) CE\left(l, q, \frac{\pi}{L}(x-y)  \right), \\
\Psi_{kl}^{-}(x,y) &= N_l^- \exp\left(\frac{\imagi \pi}{L} k (x+y) \right) SE\left(l, q, \frac{\pi}{L}(x-y)  \right),
\end{align*}
where $+$ and $-$ refer to the singlet and triplet cases respectively and $N_l^{\pm}$ is a normalization factor.
In both cases $k$ and $l$ need to be both even or both odd.
The associated energy eigenvalues are
\begin{equation*}
E^\pm_{kl}=  \left(\frac{\pi}{L}\right)^2 \left[ k^2 + a^\pm_l(q)   + 2 q \right] ,
\end{equation*}
where $a^\pm_l(q)$ are the characteristic values for the Mathieu-cosine and Mathieu-sine function respectively \cite{Stegun}.
For $q \neq 0$ the characteristic values obey $a^+_0(q)<a^-_1(q)<a^+_1(q)<a^-_2(q)<...$ , while 
in the non-interacting case $a_l^+(0)=a_l^-(0)=l^2$.  We thus nicely see how the two-particle interaction splits the degeneracy of the spin-singlet and spin-triplet states. In this noninteracting limit the wave functions attain the simple orbital product form
\begin{equation}
\Psi_{kl}^{\pm} (x,y) = N_l^{\pm} (\phi_{k+l}(x) \phi_{k-l} (y) \pm \phi_{k-l}(x) \phi_{k+l} (y))
\label{nonint}
\end{equation}
where $\phi_{n} (x)=e^{i n\pi x /L}$ and where $k \pm l$ is always even.
For any interaction strength the ground state of the QR is the spin-singlet state $\Psi_{00}^{+}(x,y)$.
We see from Eq.~(\ref{nonint}) that all states with $|k| \neq l$ correspond to doubly excited states relative to the ground state which are
notoriously difficult to describe by adiabatic functionals. We will return to this issue in Sec.~\ref{sec:freq}.
For large values of $q$ the Mathieu functions become localized around $z=\pi/2$ (and hence $r=L/2$) corresponding to the strongly correlated limit of well-localized
electrons on opposite parts of the ring. The limit $L \rightarrow \infty$ corresponds to $q \rightarrow \infty$ and to a limit where the density goes to zero.
This limit corresponds to the famous Wigner crystal \cite{GiovanniBook}.


\subsection{Exchange-Correlation Potential} 
\label{sec:vxc}

\begin{figure}
\includegraphics[width=0.49\textwidth]{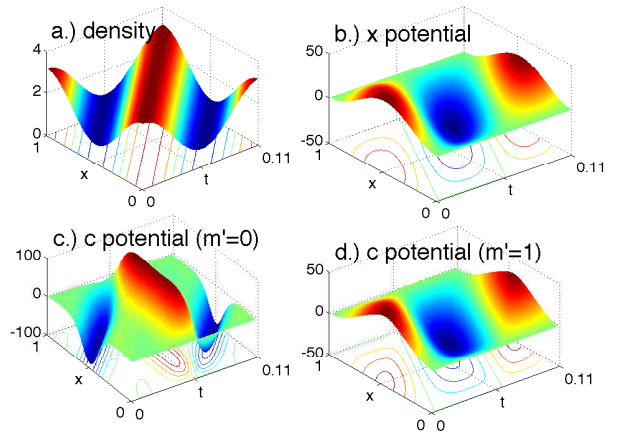} 
\caption{(color online). The density, x potential and c potentials for $C_0^2=0.5$ and $m'=0$ as well as $m'=1$ ($\lambda=100$, $L=1$). Note the change of scale between $m'=0$ and $m'=1$.
Further note, that we used the gauge-freedom of the potentials in order to set them to zero at $x=0$.}
\label{fig:3}
\end{figure}

We now start to construct the exact xc potential for a specific density $n$ that corresponds to a solution of the time-dependent Schr\"odinger equation with the Hamiltonian of Eq.~(\ref{ham1}). For such a density the xc potential is given by Eq.~(\ref{SpecificXCpot}).
The xc potential can be further split into an exchange (x) and a correlation (c) part $v_\mathrm{xc}=v_\mathrm{x}+v_\mathrm{c}$
where, for our two-electron system, the x potential is simply given by \cite{CarstenBook}
\begin{equation} 
\label{EXX}
v_\mathrm{x}([n],x,t)=-\tfrac{1}{2}v_\mathrm{H}([n],x,t).
\end{equation}
We choose the density $n$ to come from a freely propagating superposition of two normalized eigenstates of our QR
\begin{align*}
\Psi(x,y,t) &= C_0 \Psi_{00}^{+}(x,y)\exp(-\imagi E_{00}^{+}(t-t_0)) \\
&+ C_1 \Psi_{11}^{+}(x,y)\exp(-\imagi E_{11}^{+}(t-t_0)) ,
\end{align*}
which is a solution to the time-dependent Schr\"odinger equation. This wave function is properly normalized whenever
$C_0^2+C_1^2=1$. Note that both eigenstates have a constant density. If the constant $C_0$ is almost 1 (or 0), the density of the system only deviates 
slightly from being homogeneous. If we look at a small QR, e.g. $L=1$ and different interaction strengths $\lambda$, we find that even for small deviations from homogeneity the c potential is at least of the same order of magnitude as the x potential. In this case, increasing the density variations by changing $C_0$ makes the correlation potential $v_\mathrm{c}$ the dominant contribution to $v_\mathrm{xc}$.
A notable exception is an initial KS state that has approximately the right initial angular momentum
(in the case of $\lambda=100$ and $L=1$ this is the state $m'=1$ as can be seen in Fig. \ref{fig:3}).
For this case the c potential plus the x potential mainly needs to cancel the Hartree potential.
The KS orbital would travel around the ring in approximately the right manner if there were no external perturbations.
Besides the initial-state dependence one also clearly sees the non-locality of the c potential in time (memory) and space,	
as it has in general no obvious simple relation to the local density (see the c potential for $m'=0$ in Fig. \ref{fig:3}).
If we go to larger QRs, e.g. $L=2 \pi$,  the x potential becomes the dominant contribution to the xc potential.
This seems counterintuitive since for this case the value of $q$ is larger, corresponding to a more correlated state.
It should, however, be remembered that the relation between the density profile (and hence the shape of $v_s$)
and the electronic correlations is rather indirect.
For example, the ground state density and KS potential of the QR are spatially constant,
independent of the interaction strength.
To get more insight into the influence of interactions, it is therefore more useful to study a two-point function.
We will therefore now  construct the (equilibrium) xc kernel for this problem,
which is defined to be the first functional derivative of $v_\mathrm{xc}$ with respect to the density $n$,
evaluated at the ground state density.
We will be able to do so because the ground-state density of the system is homogeneous irrespective of the interaction. Therefore the $\lambda=0$ case is the KS system for any interaction strength $\lambda$.


\subsection{Exchange-Correlation Kernel} 
\label{HxcKernel}

The xc kernel is the central object of interest in linear-response TDDFT from which one can determine the perturbative dynamics of the quantum system and its excitation energies. We start by calculating how the ground state spin-density reacts to small external perturbations, i.e. 
\begin{equation} 
\label{variation}
\delta n(x \sigma, \omega) =  \sum_{\sigma'}\int \diff x' \chi (x \sigma,x' \sigma',\omega) \delta v( x' \sigma', \omega)
\end{equation}
(see e.g. in Refs.~\cite{CarstenBook,TDDFT}), where
\begin{align*}
\chi (x\sigma,x'\sigma',\omega) = \sum_{kl,p=\pm } \left[  \frac{\bra \Psi_0|\hat{n}(x\sigma)|\Psi^p_{kl} \ket \bra \Psi^p_{kl}| \hat{n}(x' \sigma')|\Psi_0 \ket}{\omega - (E^{p}_{kl}-E_0) + \imagi \epsilon} \right.
\\
-\left. \frac{\bra \Psi_0|\hat{n}(x' \sigma')|\Psi^p_{kl} \ket \bra \Psi^p_{kl}|\hat{n}(x \sigma)|\Psi_0 \ket}{\omega + (E^{p}_{kl}-E_0) + \imagi \epsilon}  \right]
\end{align*}
with $\epsilon>0$ an infinitesimal, $\hat{n}(x \sigma)$ the usual spin-density operator and $-\infty \leq k \leq \infty$ and $0 \leq l \leq \infty$ ($k$ and $l$ are always either both even or both odd).
Here with $p=-$ we refer to the triplet state with spin function
$(\delta_{\sigma,\uparrow} \delta_{\sigma',\downarrow} + \delta_{\sigma',\uparrow} \delta_{\sigma,\downarrow})/\sqrt{2}$ 
only, since the spin-triplet functions orthogonal to this one give a zero contribution in the sum.
In a first step we can deduce using the periodicity of the solutions that
\begin{align*}
\bra \Psi_0| \hat{n}(x \sigma)|\Psi^{+}_{kl} \ket &= \exp\left(\imagi 2 \pi k x /L  \right) D^{+}(k,l) ,
\\
\bra \Psi_0| \hat{n}(x \sigma)|\Psi^{-}_{kl} \ket &= \exp\left(\imagi 2 \pi k x /L  \right) D^{-}(k,l)(\delta_{\uparrow \sigma} - \delta_{\downarrow \sigma}) ,
\end{align*}
where
\begin{align*}
D^{+}(k,l)= N_0^{+} N_l^{+} \int_0^L \diff r \, &CE\left(0,q, \frac{\pi}{L} r \right ) CE\left(l,q, \frac{\pi}{L} r \right ) \\
&\times \exp\left(-\imagi \frac{\pi}{L}  k r\right) ,
\\
D^{-}(k,l)= N_0^{+} N_l^{-} \int_0^L \diff r \, &CE\left(0,q, \frac{\pi}{L} r \right ) SE\left(l,q, \frac{\pi}{L} r \right ) \\
&\times \exp\left(-\imagi \frac{\pi}{L}  k r\right) .
\end{align*}
We note that the Mathieu-cosine and Mathieu-sine are real and thus we have $D^{\pm}(k,l)^* = D^{\pm}(-k,l)$. Further we note that $D^{\pm}(0,l) = 0$ for $l \neq 0$. After some manipulations of the general expression for the linear-response kernel we end up with
\begin{align} 
\label{responsekernel} 
\chi (x \sigma,x' &\sigma',\omega) = \sum_k \bigl\{ \mu_{k}^{+}(\omega)\delta_{\sigma \sigma'} \nonumber \\
& +  \mu_{k}^{-}(\omega)\left[1- \delta_{\sigma \sigma'}\right]  \bigr\} \zeta_k(x) \zeta_k(x')^{*},
\end{align}
where
\begin{align*}
\zeta_{k}(x) &= \frac{\exp\left(\frac{\imagi  2 \pi k x}{ L} \right)}{\sqrt{L}} , \\
\mu_k^{\pm}(\omega) &= \nu_k^{+}( \omega) \pm \nu_k^{-}( \omega) , \\
\nu_k^{\pm}(\omega) &=  \sum_l \frac{ 2L \left( E_{kl}^{\pm} - E_0 \right)|D^{\pm}(k,l)|^2}{ (\omega + \imagi \epsilon)^2- \left(E_{kl}^{\pm} - E_0 \right)^2} ,
\end{align*}
where the sum runs over all even values of $l$ if $k$ is even and over all odd values if $k$ is odd.
In the non-interacting case we find due to
$|D^{\pm}(k,l)|^2 \rightarrow \delta_{|k|,l}/(2L^2)$ and $\nu_k^{0,+}(\omega) = \nu_k^{0,-}(\omega)$ 
the simple expressions
\begin{equation*}	
\mu_k^{0,+}(\omega) = 2 \nu_k^{0,+}(\omega) 
= \frac{1}{L}\left( \frac{2 \pi}{L}\right)^2 \frac{ k^2}{ (\omega+\imagi \epsilon)^2-\frac{1}{4}\left(\frac{2 \pi}{L}\right)^4 k^4 \ }
\end{equation*}
and $\mu_k^{0,-}(\omega) = 0$. Thus the non-interacting linear response kernel $\chi_0$ has non-zero contributions only from excited states with $|k|=l$. As discussed below Eq.~(\ref{nonint}) the states with $|k|=l$ are exactly the singly-excited states of the non-interacting system. We therefore recovered the well-known fact that the non-interacting response function $\chi_0$ has only poles at singly-excited states.

In linear-response (spin) TDDFT the interacting response function $\chi$ is expressed in terms of the response function of a non-interacting system with the same density. In our case, since the ground-state density is homogeneous irrespective of the interaction strength $\lambda$, the KS system is the one with $\lambda=0$ and the corresponding KS response function is $\chi_0$. Therefore we can express 
\begin{equation} 
\label{f_xc}
\chi = \chi_0 + \chi_{0} f_\mathrm{Hxc} \chi ,
\end{equation}
where the Hartree-exchange-correlation (Hxc) kernel is defined as
\begin{equation*}
f_\mathrm{Hxc} = \chi_0^{-1}-\chi^{-1} ,
\end{equation*}
and integration as well as summation over reoccurring position-spin variables is implied. With the inverse
kernels of Eq.~(\ref{responsekernel}) we find that
\begin{multline*}
f_\mathrm{Hxc}(x \sigma,x' \sigma', \omega)
\\
= \sum_{k \neq 0} \biggl\{  \left(\frac{1}{\mu^{0,+}_k(\omega)} - \frac{\mu_{k}^{+}(\omega)}{ 4 \nu_{k}^{+} (\omega)\nu_{k}^{-}( \omega)} \right) \delta_{\sigma \sigma'} 
\\
+  \frac{\mu_{k}^{-}(\omega)}{ 4 \nu_{k}^{+}( \omega) \nu_{k}^{-}( \omega)} \left[1- \delta_{\sigma \sigma'}\right]  \biggr\} \zeta_k(x) \zeta_k(x')^{*}.
\end{multline*}
The xc kernel is then trivially found by subtracting the interaction potential, i.e.  $f_\mathrm{xc} = f_\mathrm{Hxc} - w$. 
The restriction to $k \neq 0$ in the sum is a consequence of the fact that the response functions are only invertible in the space of functions orthogonal
to constant function, since a constant potential variation gives no density change. For the Hxc kernel this amounts to the freedom of adding any function of the form $g(x\sigma,x'\sigma',\omega) = g_1(x\sigma,\omega) + g_2(x'\sigma',\omega)$, since it is always constant either in $x'\sigma'$ or $x\sigma$ when integrating over the internal degrees of freedom in Eq.~(\ref{f_xc}). Therefore adding a function $g$ to $f_\mathrm{Hxc}$ does not change the linear response kernel $\chi$ \cite{Hellgren2012}. 
We have now fully characterized the behavior of the interacting particles on a QR in terms of the KS system for weak external perturbations. The xc kernel exhibits a strong frequency dependence as it needs to shift the poles of $\chi_0$ and generate new poles in order to have the correct density response of the correlated system.
If we Fourier-transformed the kernel from frequency to time, the frequency-dependence would translate to a dependence on previous times, i.e. the frequency-dependence corresponds to memory. Therefore we have constructed the first exact density-functional with memory.


\subsection{Frequency-Dependence of the Exchange-Correlation Kernel}
\label{sec:freq}

\begin{figure}[H]
\includegraphics[width=0.49\textwidth]{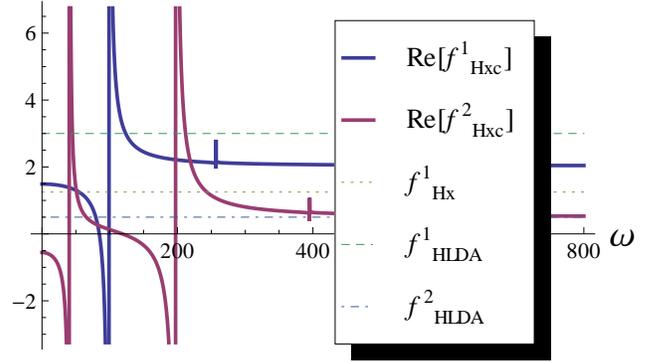} 
\caption{(color online). The real part of $f_\mathrm{Hxc}^k$, $f_\mathrm{Hx}^k$ and $f_\mathrm{HLDA}^k$ for $k=1,2$ ($\lambda=10$ and $L=1$).}
\label{fig:kernel}
\end{figure}

In a next step we investigate the frequency dependence of the Hxc kernel in more detail.
Such considerations are of importance for developing frequency-dependent approximations to the Hxc kernel \cite{Gross1985,Ullrich1995,Vignale1996,Sottile2003,GoerlingBook,Hellgren2010},
since even advanced approximation schemes can result in unphysical behaviour \cite{Hellgren2009}.
To simplify the forthcoming discussion a little we will restrict ourselves to spin-independent linear-response theory,
i.e. we only allow for spin-independent pertubations $\delta v(x,\omega)$ in Eq.~(\ref{variation})
and are interested in $\delta n(x,\omega) = \sum_\sigma \delta n(x \sigma,\omega)$.
Therefore we can straightaway sum over all spin-degress of freedom in Eq.~(\ref{responsekernel}), leading to
\begin{equation*}
\chi(x,x',\omega) =  \sum_k 4 \nu_k^+(\omega) \zeta_k(x) \zeta_k^*(x') .
\end{equation*}
Accordingly we no longer couple to the spin-triplet states, and of the whole physical spectrum
\begin{equation*}
\Delta E^\pm_{kl} = \left(\frac{\pi}{L}\right)^2 \left[ k^2 + a^\pm_l(q) - a^+_0(q) \right]
\end{equation*}
only the spin-singlet transitions $\Delta E^{+}_{kl}$ show up in our linear-response calculations. If we further note that $\zeta_k(x) = \bra x|k \ket$ is a spatial basis (for square-integrable functions) we can express
\begin{align*} 
\hat{f}_\mathrm{Hxc}(\omega) &=  \sum_{k \neq 0}   |k \ket \, \frac{1}{4}\left(\frac{1}{\nu^{0,+}_k(\omega)} - \frac{1}{ \nu_{k}^{+} (\omega)} \right) \bra k| \\
&= \sum_{k \neq 0}   |k \ket \, f^k_\mathrm{Hxc}(\omega) \, \bra k| . \nonumber
\end{align*}
It is now interesting to compare the exact expression to some standard approximations for the Hxc kernel.  We first note that
\begin{equation*}
w(x-x')=\frac{\lambda L}{4} \left( \bra x |-1 \ket \bra -1|x' \ket + \bra x|1 \ket \bra 1|x' \ket + \frac{2}{L}\right).
\end{equation*}
Therefore the Hartree-exchange approximation (Hx) reads with $\hat{\mathbb{1}} = \sum_k |k \ket  \bra k|$
\begin{equation*}
\hat{f}_\mathrm{Hx}(\omega) = \sum_{k = \pm 1}   |k \ket \, \frac{\lambda L}{8} \bra k| + \frac{\lambda}{4} \, \hat{\mathbb{1}},
\end{equation*}
since according to Eq.~(\ref{EXX}) it is simply obtained by functional differentiation $1/2$ of the Hartree term for the case of a two-particle spin-singlet 
state. The local-density approximation (LDA) together with the Hartree (H) term amounts to
\begin{equation*}
\hat{f}_\mathrm{HLDA}(\omega) =  \sum_{k \neq 0} |k \ket \, \epsilon''_{\mathrm{QR}} \bra k| + \sum_{k = \pm 1}   |k \ket \, \frac{\lambda L}{4} \bra k| + \frac{\lambda}{2} \, \hat{\mathbb{1}} ,
\end{equation*}
where $\epsilon''_{\mathrm{QR}}$ is determined by the second functional derivative of the xc energy functional of the (homogeneous) ground-state density \cite{CarstenBook}, i.e.,
\begin{equation*}
\left.\frac{\delta^2 E_\mathrm{xc}^{\mathrm{LDA}}[n]}{\delta n(x) \delta n(x')}\right|_{n=n_0} = \epsilon''_{\mathrm{QR}} \delta(x-x').
\end{equation*}
We approximate it from Fig.~\ref{fig:kernel}, where we employ a value of $\epsilon''_{\mathrm{QR}} \simeq 0.5$ such that we on average reproduce the exact $f_\mathrm{Hxc}^k = \langle k | \hat{f}_{\mathrm{Hxc}}| k \rangle$ for $|k|>1$. We compare the exact expression for the Hxc kernel to the (frequency-independent) approximations for $k=1$ and $2$ in Fig. \ref{fig:kernel}. The Hx approximation has only a contribution for $k=1$ while the HLDA approximation has a contribution for every value of $k$. We further see in Fig.~\ref{fig:kernel} that for $|k|>1$ and $\omega \rightarrow \infty$ the HLDA approximation and the exact kernel become identical.

In order to understand how the frequency-dependence that is missing in the above approximations works,
it is useful to express the interacting kernel $\hat{\chi}$ in a different form. From Eq.~(\ref{f_xc}) we find that
\begin{equation*}
\hat{\chi} = \frac{\hat{\chi}_0}{\hat{\mathbb{1}} - \hat{\chi}_0 \hat{f}_\mathrm{Hxc}} .
\end{equation*}
The task of the denominator $\hat{\mathbb{1}}-\hat{\chi}_0 \hat{ f}_\mathrm{Hxc}$ is two-fold: it shifts the existing poles of $\hat{\chi}_0$ and it generates poles that are missing in the bare KS kernel. In order to do so, the denominator has to become zero at the values of the physical resonance frequencies $\Delta E^+_{kl}$. This condition reads as
\begin{equation*}	
\sum_{k'} |k'\ket  \bra k'| = \sum_{k', k''} |k' \ket \bra k'| \hat{\chi}_0 \hat{f}_\mathrm{Hxc}(\Delta E^\pm_{kl})|k''\ket \bra k''| .
\end{equation*}
Therefore $\bra k'| \hat{\chi}_0 \hat{f}_\mathrm{Hxc} (\Delta E^\pm_{kl})|k''\ket = \delta_{k'k''}$. If we interpret $\hat{\chi}_0 \hat{f}_\mathrm{Hxc}$ as an infinite-dimensional matrix (in the above basis set), then at the resonance frequencies it has only entries in the diagonal and is zero otherwise. We see that in our case the matrix expression of $\hat{f}_\mathrm{Hxc}$ is already diagonal for any frequency. This does also not change if we multiply by the matrix expression for $\hat{\chi}_0$ and find
\begin{equation*}
\hat{\mathbb{1}} - \left(\hat{\chi}_0 \hat{f}_\mathrm{Hxc}\right)(\omega) = \sum_{k \neq 0}   |k \ket \left( \frac{\nu^{0,+}_k(\omega)}{\nu_k^+(\omega)}\right) \bra k| .
\end{equation*}
We immediately see that when $\hat{\chi}$ has a pole ($\nu^+_k(\omega) \rightarrow \infty$) then $\nu^{0,+}_k(\omega)/ \nu_{k}^{+}(\omega) \rightarrow 0$. 
\begin{figure}[H]
\includegraphics[width=0.49\textwidth]{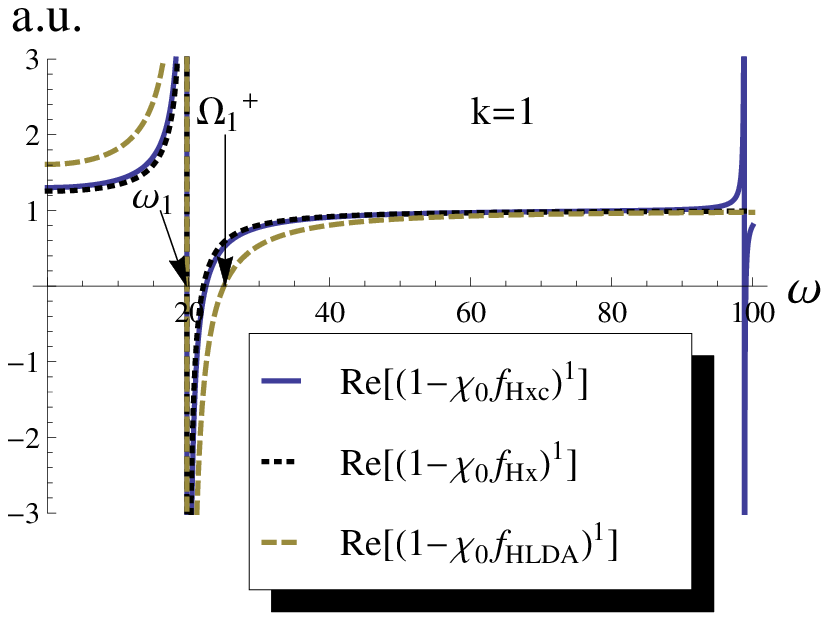} 
\includegraphics[width=0.49\textwidth]{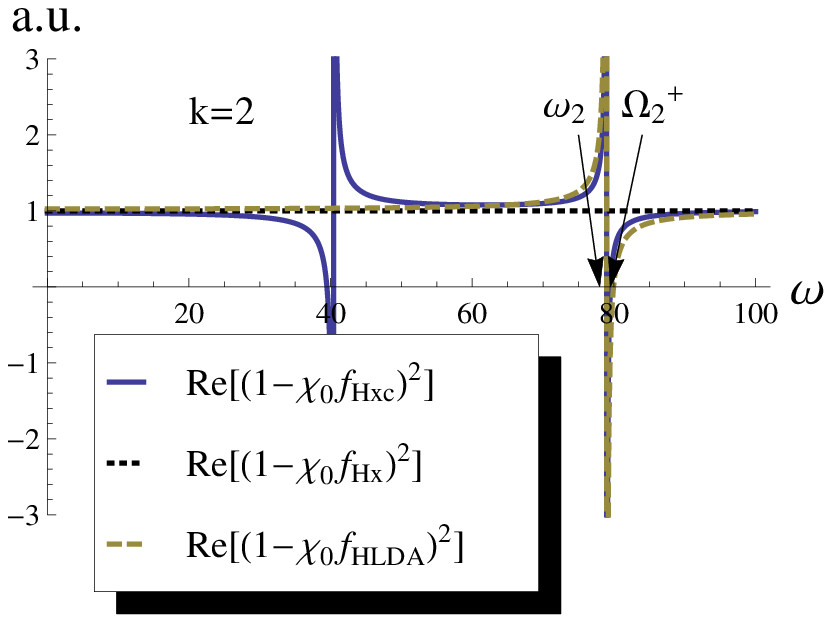} 
\caption{(color online). The real parts of $(1 - \chi_0 f_\mathrm{Hxc})^k$, $(1 - \chi_0 f_\mathrm{Hx})^k$ and $(1 - \chi_0 f_\mathrm{HLDA})^k$ for $k=1,2$ ($\lambda=10$ and $L=1$). The bare KS resonances $\omega_1$ and $\omega_2$ are indicated with arrows pointing to their values on the frequency axis. We have also indicated the single-pole approximated resonances $\Omega^{+}_{1}$ and $\Omega^{+}_{2}$ (in this frequency range) by arrows pointing to their respective values.}
\label{fig:inverse1}
\end{figure}
This behaviour is nicely visible in Fig.~\ref{fig:inverse1}, where for $k=1$ and $k=2$ we have zeros at the first four eigenfrequencies of the interacting system ($\Delta E_{11}^{+} = 22.5$, $\Delta E_{20}^{+} = 39.5$, $\Delta E_{22}^{+} = 79.5$ and $\Delta E_{13}^{+} = 99.0$). 
As explained below Eq.~(\ref{nonint}) the excitations of the form $\Delta E_{(\pm l) l}^{+}$ are singly-excited states, whereas the excitations of the
form $\Delta E_{kl}^{+}$ with $|k| \neq l$ correspond to doubly-excited states which do not generate poles in the noninteracting response function.
Indeed, we see that the Hxc kernel generates new poles ($\Delta E_{20}^{+}$ and $\Delta E_{13}^{+}$) in the exact response function corresponding to doubly-excited states which are missing entirely in the KS kernel $\hat{\chi}_0$. In contrast, the frequency-independent approximations can only shift the already existing poles. The exact kernel does this by canceling the bare KS poles since $\nu^{0,+}_k(\omega) \rightarrow \infty$, and generates the corresponding physical poles ($\Delta E_{11}^{+}$ and $\Delta E_{22}^{+}$). We note that the Hx approximation does only change the position of the first KS resonance but leaves all others unmodified.


\subsection{Single-Pole Approximation}
\label{sec:SPA}

\begin{figure}[H]
\includegraphics[width=0.49\textwidth]{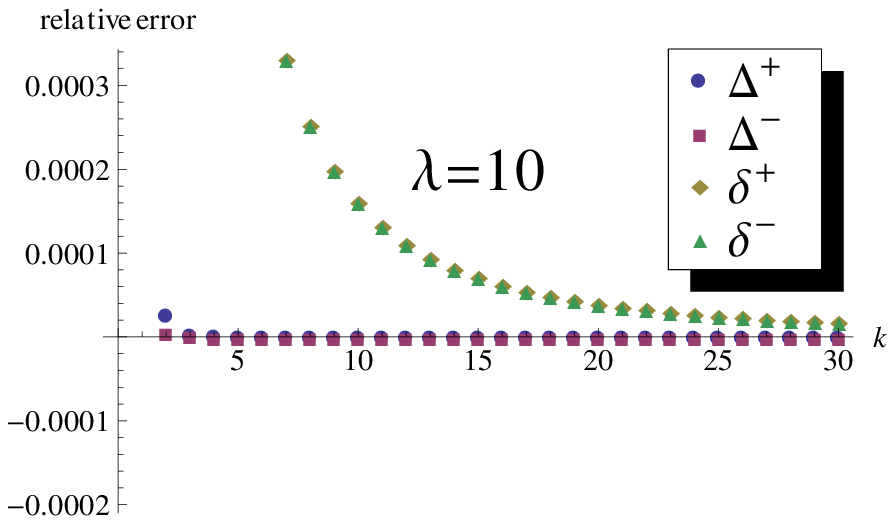}
\caption{(color online). The relative error $\Delta^{\pm}_k$ of the SPA and the relative error $\delta^{\pm}_k$ of the bare KS resonances for $\lambda=10$ and $L=1$.}
\label{fig:lambda10}
\end{figure}

\begin{figure}[H]
\includegraphics[width=0.49\textwidth]{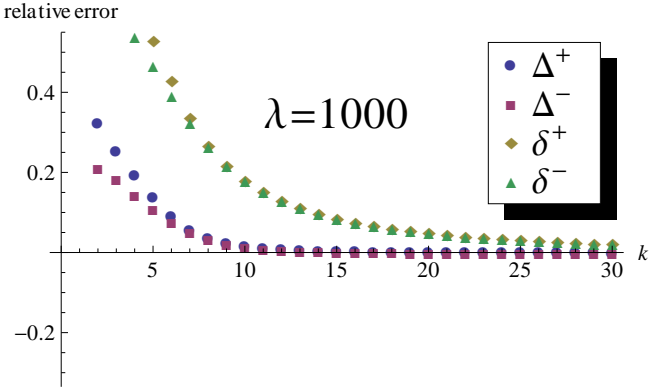}
\caption{(color online).  The relative error $\Delta^{\pm}_k$ of the SPA and the relative error $\delta^{\pm}_k$ of the bare KS resonances for $\lambda=1000$ and $L=1$.}
\label{fig:lambda1000}
\end{figure}

In practice, even if one had the exact Hxc kernel, calculations are often performed employing certain approximations \cite{CarstenBook, TDDFT}. One of the most important approximations used to determine excitation energies from a linear-response TDDFT calculation is the so-called single-pole approximation (SPA) \cite{CarstenBook, TDDFT}. This approximation can be derived from our previous considerations on the Hxc kernel, from where we know that if $\hat{\mathbb{1}}-\hat{\chi}_0 \hat{f}_\mathrm{Hxc} = 0$ for some frequency $\omega$ then it corresponds to a physical resonance. This equation can be rewritten in terms of a generalized eigenvalue problem called the Casida equation \cite{Casida1995}. By expanding in terms of KS frequencies $\omega_k$ \cite{Petersilka1996} one finds that the spin-singlet and spin-triplet excitation energies are perturbatively given by
\begin{equation*}
\Omega_k^\pm = \omega_k + 2 \Re\left\{\int \diff x \int \diff x' \Phi_{0k}^*(x) k^\pm(x,x',\omega_k) \Phi_{0k}(x)   \right\}
\end{equation*}
where $ \Phi_{0k}(x)= \varphi_0(x) \varphi_k(x)$ and $\varphi_k(x)$ the $k$-th KS orbital as well as
\begin{align*}
k^{+}(x,x',\omega_k) = &\frac{1}{2}\!\left[ f_\mathrm{xc}(x \!\! \uparrow,x'\!\!\uparrow,\omega_k) +  f_\mathrm{xc}(x \!\! \uparrow,x'\!\! \downarrow,\omega_k) \right] , \\
&+ w(x-x'), \\
k^{-}(x,x',\omega_k) = &\frac{1}{2}\!\left[ f_\mathrm{xc}(x \!\! \uparrow,x'\!\!\uparrow,\omega_k) -  f_\mathrm{xc}(x \!\! \uparrow,x'\!\! \downarrow,\omega_k) \right].
\end{align*}
In the case at hand we can perform these integrals analytically and find by using that $1/\mu^{0,+}_k (\omega_k) = 0$ the simple expression
\begin{equation}
\label{SPA}
\Omega_k^\pm = \omega_k -\frac{1}{L} \Re\left\{\frac{1}{2 \nu^{\pm}_k(\omega_k)}\right\} \pm \frac{\lambda}{4} \, \delta_{|k|,1} .
\end{equation}
In Fig.~\ref{fig:inverse1} we have indicated the first two shifted (singlet-singlet) eigenvalues calculated with the SPA ($\Omega^{+}_1 = 25.2$ and $\Omega^{+}_2 = 79.5$). While the first shift of the eigenvalues is overestimated (from the bare KS resonance $\omega_1 = 19.7$) the second resonance frequency is extremely well reproduced (with the bare KS resonance being $\omega_2 = 78.9$). Still the calculation of the SPA results includes a sum over (infinitely many) $l$ values. In order to more easily investigate the behaviour of the SPA for a large set of resonances and cases we make a further approximation to find a closed expression for $\Omega_k^{\pm}$. To do so we note, that the term $|D^{\pm}(k,l)|^2$ gives its main contribution for $k=l$ and falls off rapidly. Therefore it seems a reasonable approximation to employ the value of the non-interacting case, i.e. $|D^{\pm}(k,l)|^2 \rightarrow \delta_{|k|,l}/(2L^2)$. This leads to the simple explicit expression
\begin{equation} \label{SPAsimple}
\Omega_k^\pm \simeq \omega_k  + \frac{1}{2} \frac{(\Delta E_{kk}^\pm)^2- \omega_k^2}{\Delta E_{kk}^\pm} \pm \frac{\lambda}{4} \, \delta_{|k|,1} .
\end{equation}
With this explicit expression \footnote{If one compares the approximation of Eq.~(\ref{SPAsimple}) with the exact SPA of Eq.~(\ref{SPA}) for some lower lying resonances, the approximation seems to perform usually better, at least if the summation over $l$ is truncated at some finite value.} we can now easily investigate properties of the SPA. We do so by comparing the relative error of the SPA
\begin{equation*}
\Delta^\pm_k = (\Delta E_{kk}^{\pm} -\Omega^{\pm}_k)/\Delta E_{kk}^{\pm}
\end{equation*}
with the relative error of the bare KS eigenvalues
\begin{equation*}
\delta^\pm_k = (\Delta E_{kk}^{\pm} -\omega_k)/\Delta E_{kk}^{\pm}.
\end{equation*}
The results are displayed in Figs.~\ref{fig:lambda10} and \ref{fig:lambda1000}.
We first note, that the bare KS response does not have a spin-triplet component. Nevertheless, we take the bare KS resonances as a zeroth order guess for both, the spin-singlet and spin-triplet transitions. Therefore in general $\omega_k$ has a different relative error with respect to $\Delta E_{kk}^{\pm}$. This can be nicely seen in Fig. \ref{fig:lambda1000} in the difference between $\delta^+_k$ and $\delta^-_k$ for small values of $k$. For large values of $k$ these differences as well as the relative error of the different approximations becomes small. For the bare KS resonance frequencies this can be explained by the fact, that the higher lying states are dominated by the kinetic energy and are less influenced by the interaction. On the other hand, since the SPA corresponds to the first term in a Laurent expansion with respect to $\omega_k$ \cite{Petersilka1996}, it becomes more accurate when the bare KS excitation energy is closer to the true resonance frequency. Hence the SPA inherits the high $k$ behaviour of the bare KS values. For the lower lying states we find that the SPA usually strongly improves upon the bare KS resonances (see Figs.~\ref{fig:lambda10} and \ref{fig:lambda1000}). In our example the singly-excited states are well-separated
from the doubly-excited states. In this case the SPA describes well the singly-excited states. 

The adiabatic approximations based on LDA and the 
exchange-only kernel do not generate new poles and, not surprisingly, fail to describe the doubly-excited states of the system. In order to describe
doubly-excited states a frequency-dependent xc kernel is required. We are not aware of any existing simple approximation for a memory kernel that
would be able to reproduce the doubly-excited states of our model system. Some approximate kernels exist based on electron gas models, but such
kernels would lead to artificial complex excitation energies. In any case, our exact expression for the xc kernel will serve as a useful benchmark
to study future density functionals.


\section{Outlook}
\label{sec:Conclusion}
					
In this work we have presented analytical expressions of exact density-functionals with initial-state dependence and memory. The functionals were used to give explicit examples of the otherwise very abstract concepts of the xc potential and the KS construction. We demonstrated how one can calculate the exact xc potential for an interacting model system and how one can construct the corresponding exact, frequency-dependent xc kernel. We have then shown how these analytical examples can be used to investigate the basic properties of time-dependent density-functionals and to test approximations. 

These results will help to understand the properties of time-dependent density-functionals in more detail.
For instance, they have already been used to investigate the Floquet-approach to TDDFT \cite{Kapoor2013}.
Further, these exact functionals show how initial-state dependence and memory have to be incorporated into more accurate functional approximations. As such one can employ these exact expressions as benchmarks for the development of new and more reliable functional approximations in TDDFT.

\begin{acknowledgments}
We thank E.\ R\"as\"anen, N.T.\ Maitra and K.\ Burke for valuable comments.
M.R. acknowledges financial support by the Erwin Schr\"odinger Fellowship J 3016-N16 of the FWF (Austrian Science Fund). 
S.E.B.N. acknowledges support from the Lundbeck Foundation and from a grant to prof. Jeppe Olsen from the Danish Research Council. R.v.L. acknowledges the Academy of Finland for research funding.
\end{acknowledgments}

\bibliography{Library}


\end{document}